# Deterministic Spin-Orbit Torque Switching of Mn$_3$Sn with the Interplay between Spin Polarization and Kagome Plane

Zhengde Xu[1], Xue Zhang[1], Yixiao Qiao[1], Gengchiau Liang[2,3], Shuyuan Shi[4†] and Zhifeng Zhu[1,5†]

[1]School of Information Science and Technology, ShanghaiTech University, Shanghai, China 201210

[2]Department of Electrical and Computer Engineering, National University of Singapore, Singapore 117576

[3]Industry Academia Innovation School, National Yang-Ming Chiao Tung University, Hsinchu City, Taiwan 300093

[4]Fert Beijing Institute, MIIT Key Laboratory of Spintronics, School of Integrated Circuit Science and Engineering, Beihang University, Beijing 100191, China

[5]Shanghai Engineering Research Center of Energy Efficient and Custom AI IC, Shanghai, China 201210

**Abstract**

Previous studies have demonstrated spin-orbit torque (SOT) switching of Mn$_3$Sn where the spin polarization lies in the kagome plane (configuration I). However, the critical current density ($J_{\text{crit}}$) is unrealistically large ($J_{\text{crit}}=10^{14}$ A/m$^2$) and independent on the external field ($H_{\text{ext}}$). The stabilized magnetic state also depends on the initial state. These features conflict with the ferromagnet (FM) switching scheme as claimed in those studies, and thus call for other explanations. Alternatively, the system with the spin polarization perpendicular to the kagome plane (configuration II) is more like the FM based system since the spin polarization is orthogonal to all magnetic moments. In this work, we show SOT switching of Mn$_3$Sn in configuration II. Similar to the FM, $J_{\text{crit}}$ and $H_{\text{ext}}$ are in the order of $10^{10}$ A/m$^2$ and hundreds of Oersted, respectively.

The switching result is also independent of the initial state. Interestingly, the unique spin structure of Mn$_3$Sn also leads to distinct features from FM systems. We demonstrate that $J_{crit}$ increases linearly with $H_{ext}$, and extrapolation gives ultralow $J_{crit}$ for the field-free switching system. In addition, the switching polarity is opposite to the FM. We also provide the switching phase diagram as a guideline for experimental demonstration. Our work provides comprehensive understanding for the switching mechanism in both configurations. The switching protocol proposed in this work is more advantageous in realistic spintronic applications. We also clearly reveal the fundamental difference between FM and noncollinear antiferromagnetic switching.

**Introduction**

The increasing demand for higher memory performance stimulates the research of magnetic random access memory (MRAM). In the last two decades, the development of current control of magnetic states in ferromagnets (FM) [1,2] enables fast switching in sub-nanoseconds [3], low switching current of $10^9$ A/m$^2$ [4], and small device in nanometer diameters [5]. To further improve the performance, in recent years people turned their attention to antiferromagnets (AFM) which consist of antiparallelly aligned magnetic moments. They can be roughly divided into collinear AFM and noncollinear AFM. Compared to the widely studied collinear AFM [6-9], the spin-orbit torque (SOT) switching of noncollinear AFM remain to be explored.

Mn$_3$Sn is an exemplary material for the noncollinear AFM [10-14]. It has exotic properties such as the large anomalous Hall effect (AHE) [15]. This enables effective read out of the magnetic state. The SOT switching of Mn$_3$Sn was first demonstrated in a bilayer Mn$_3$Sn/Heavy Metal (HM) structure, in which the spin polarization lies in the kagome plane (configuration I) [16-18]. The authors attribute the switching mechanism the same as FM. However, the high critical switching current ($J_{crit}=10^{14}$ A/m$^2$) required in the simulation does not match that in experiments, which has motivated further studies that attribute the much smaller $J_{crit}$ from experiments to the indispensable role from the thermal effect [19,20]. In fact, there can be three configurations in the SOT switching of Mn$_3$Sn. As illustrated in Fig. 1(c), in configuration II, the spin polarization is perpendicular to the kagome plane, whereas in configuration III the kagome plane is parallel to the thin film. In configuration II, SOT induced magnetization oscillation has been observed [16,21]. A recent study even provided experimental evidence for the magnetization switching [22], but it was attributed to the misalignment of crystalline orientation and thus classified to configuration I. Intuitively, configuration II is more like the SOT switching of perpendicular FM since the spin polarization **σ** is orthogonal to all magnetic moments. Therefore, it is crucial to explore and understand the switching behaviors in configuration II.

In this work, we first provide a complete magnetization dynamics diagram for configuration I. Besides reproducing the large $J_{crit}$ [16,19], we find the switching results are sensitive to the initial state. In addition, $J_{crit}$ is independent of $H_{ext}$ [16,19,22]. The configuration I is therefore very different from the FM switching. We then study the magnetization dynamics in configuration II. We first reproduce the magnetization oscillation when **H**$_{ext}$ is absent or small. Interestingly, further increase in **H**$_{ext}$ leads to successful switching, with $J_{crit}$ in the order of $10^{10}$ A/m$^2$ and $H_{ext}$ of 100 Oe. In addition, we show $J_{crit}$ varies with $H_{ext}$, and the switching result is independent of the initial state. These features are highly similar to FM switching. However, the complex spin structure in

Mn3Sn produces many distinct features, e.g., $J_{crit}$ increases with $H_{ext}$, faster and precession less switching compared with that in FM, and opposite switching polarities to FM. Finally, we provide a phase diagram to guide future experiment verification. Our work reveals the importance of considering the spin configuration in the nuncollinear AFM switching. It also provides insights to better understand the magnetization dynamics in Mn3Sn-like AFMs.

**Methodology**

Mn3Sn is studied using the three sublattice macrospin model. As shown in Fig. 1(a), the three sublattice are marked as red circles. The energy of this system is determined by the following Hamiltonian,

$$\mathbf{H} = A\sum_{i,j}\mathbf{m}_i \cdot \mathbf{m}_j + \sum_{i,j}\mathbf{D}_{ij} \cdot (\mathbf{m}_i \times \mathbf{m}_j) - \sum_i (\mathbf{K}_i \cdot \mathbf{m}_i)^2 - \mu_0 m_s \sum_i (\mathbf{m}_i \cdot \mathbf{H}_{ext}),$$

where the exchange interaction constant $A$ = 17.53 meV, the Dzyaloshinskii–Moriya (DM) interaction $D$ = 0.833 meV, the magnetic anisotropy constant $K$ = 0.196 meV, and the magnetic moment $m_s = 3\mu_B$. All these parameters are the same as [13]. We only consider the exchange and DM interaction between the nearest neighbors in the same layer [find a reference]. The easy axis of crystalline anisotropy is marked by the yellow arrows, which point to the nearest Sn atoms. The last term is the Zeeman energy due to the external magnetic field $\mathbf{H}_{ext}$.

The dynamics of magnetic moments are described by the coupled Landau–Lifshitz–Gilbert–Slonczewski (LLGS) equations [23-25],

$$\frac{d\mathbf{m}_i}{dt} = -\gamma \mathbf{m}_i \times \mathbf{H}_{eff,i} + \alpha \mathbf{m}_i \times \frac{d\mathbf{m}_i}{dt} - \gamma \frac{\hbar \theta_{SH} J_c}{2eM_s d} \mathbf{m}_i \times (\mathbf{m}_i \times \boldsymbol{\sigma}_i).$$

The first term on the right-hand side represents the precession of the magnetic moment around the effective magnetic field $\mathbf{H}_{\text{eff},i} = -\frac{1}{m_s}\frac{\partial H}{\partial \mathbf{m}}$. The second term describes the Gilbert damping, and the last term is the SOT. $\gamma$ is the gyromagnetic ratio, and $\alpha = 0.003$ [16] is the damping constant. The spin-Hall angle $\theta_{SH} = 0.06$ [13] and the thickness of the Mn$_3$Sn film $d = 30$ nm [13] are used in the simulation. The saturation magnetization $M_s$ is calculated as $M_s = 6m_s/V_{\text{cell}}$ [16] with the unit cell volume $V_{\text{cell}} = 0.3778$ nm$^3$. The coupled LLGS equation are numerically solved via the Runge-Kutta fourth-order method. The simulation time step is 5 fs. The relaxation of the system leads to a finite net magnetization comparable to that previously reported [16].

**Results and Discussion**

In equilibrium, Mn$_3$Sn has six states that are equivalent in energy. As shown in Fig. 1(b), we define them as state a, b, c, d, e, and f, respectively. In the Mn$_3$Sn/HM bilayer, there can be three configurations where the kagome plane is perpendicular to either of the three orthogonal axis [16]. Previous studies have demonstrated SOT switching of Mn$_3$Sn in configuration I [16,19-21]. We have reproduced these results in supplementary S1, i.e., starting from state c, it can be switched to state d and vice versa [16,19]. Some studies claim the switching mechanism is the same as that in FM. However, $J_{\text{crit}}$ required in simulation is very large (i.e., in the order of 10$^{14}$ A/m$^2$ [see Fig. S1 and [16,19]] that inconsistent with experimental values. This discrepancy motivates recent studies that conclude the crucial role of thermal effect [19,20]. Besides the large current density required to switch Mn$_3$Sn in configuration I, we must notice other important differences compared to that in FM. Since the six states are equivalent, there will be in total six switching cases where each stable state is used as the initial state. Previous studies only discussed two of them [16,19], i.e., starting from states c and d. We first complete all the switching cases. As shown in table S1, states a and f can be reached by applying $\mathbf{J}_c$ and $\mathbf{H}_{ext}$, but they have no effect on state b and e. In addition, when $\mathbf{J}_c$ is along −x direction and $\mathbf{H}_{ext}$ is along +x direction, state a is changed to state f, but state c is changed to state d. This shows that under a fixed combination of $\mathbf{J}_c$ and $\mathbf{H}_{ext}$, the switching results are sensitive to the initial state. This is in stark contrast to the SOT switching of perpendicular FM, in which the direction of switching is independent of the initial state, e.g., when both $\mathbf{J}_c$ and $\mathbf{H}_{ext}$ are along +y direction, the magnetization in FM will always favor −z direction. In addition, it has been experimentally demonstrated that $J_{crit}$ is independent of $H_{ext}$ in configuration I [16,19,22], whereas $J_{crit}$ decreases with $H_{ext}$ in FM [26] . Most importantly, since $\sigma$ lies in the kagome plane in configuration I, there will be collinear component between $\sigma$ and magnetic moments, which is different from the switching of perpendicular FM where $\sigma$ is perpendicular to the magnetization. In fact, configuration II is more like the FM system since $\sigma$ is orthogonal to all magnetic moments. In addition, the corresponding effective SOT field lies in the kagome plane, resulting in an effective control. Therefore, we expect that there will be magnetization switching in configuration II.

In configuration II, when $\mathbf{H}_{ext}$ is absent, the magnetic moments develop into oscillation [Fig. S2(a)], which has been demonstrated previously [21]. The case with $\mathbf{H}_{ext}$ applied along $\mathbf{J}_c$ has been studied before [16], where they predict oscillation instead of deterministic switching. As shown in Fig. S2(b), under a small $\mathbf{H}_{ext}$ = 50 Oe, this oscillation has also been reproduced. However, when we apply $J_c$=5×10$^{10}$ A/m$^2$ and $\mathbf{H}_{ext}$ is increased to 100 Oe along +y direction, as shown in Fig. 2(a), deterministic switching is observed, i.e., state c is switched to state a. When $\mathbf{H}_{ext}$ is reversed, state c is switched to state d [see Fig. 2(b)]. When we further reverse $\mathbf{J}_c$, state c is switched to state f [see Fig. 2(c)]. These results demonstrate that both SOT and $\mathbf{H}_{ext}$ are indispensable in the switching of Mn$_3$Sn. In addition, $J_{crit}$ is in the order of 10$^{10}$ A/m$^2$. As shown in Fig. 2(d), we also verified that the switching direction only depends on the combination of $\mathbf{J}_c$ and $\mathbf{H}_{ext}$ and independent of the initial state, e.g., regardless of the initial state, $\mathbf{J}_c$ and $\mathbf{H}_{ext}$ along +y direction will always bring the system to state a. Both features are similar to FM. The complete switching cases are summarized in table 1. From the table, we noticed that states b and e can never be reached, which happens to give zero AHE signal. We also notice that the switching process has less precession, which is a key feature of AFM that possesses strong exchange coupling between sublattices. This also leads to ultrafast switching completed in sub-nanoseconds as shown in Fig. 2(a).

Since the magnetic moment in Mn$_3$Sn is more sensitive to the in-plane effective field [27], we propose to understand our switching results by analyzing the effective field that exerts on the magnetic octupole. The effective field corresponding to the damping-like torque acting on each magnetic moment is in the form $\mathbf{H}_{eff\_DLT}=\dfrac{\hbar\theta_{SH}J_c}{2eM_s d}\boldsymbol{\sigma}\times\mathbf{m}_{oct}$. As shown in Fig. 3(a), $\mathbf{J}_c$ in the +y direction gives $\boldsymbol{\sigma}$ along +x direction, resulting in $\mathbf{H}_{eff\_DLT}$ shown as the blue arrow. $\mathbf{H}_{eff\_DLT}$ compensates for $\mathbf{H}_{ext}$ which points in the +y direction. Following this principle, the other three cases can be derived as shown in Figs. 3(b)-(d). These results are consistent with those obtained from numerical simulations as shown in table 1. Now we can easily understand why states b and e can never be reached. Since $\mathbf{m}_{oct}$ in these two states points to 90° and 270°, the corresponding torque from $\mathbf{H}_{ext}$ vanishes. Therefore, the balance between effective fields cannot be achieved under a finite $\mathbf{J}_c$. These two states will finally evolve into either of the four states shown in Figs. 3(a)-3(d).

Following this explanation, one expects that $J_{crit}$ should become larger when $H_{ext}$ is increased. This has been verified in the phase diagram shown in Fig. 3(e). In the phase diagram, the initial state is state d, and the successful switching to state a is denoted as region 1 using the blue color. As $H_{ext}$ is increased from 25 Oe to 500 Oe, $J_{crit}$ increases linearly from $3.2 \times 10^{10}$ A/m$^2$ to $7 \times 10^{10}$ A/m$^2$. Noticed that the strength of $\mathbf{H}_{eff\_DLT}$ is a linear function of $J_c$, the linear trend shown in the phase diagram again supports our explanation that the switching is realized by the balance of effective SOT field and the external field. In contrast, in the FM system, it is well known that $J_{crit}$ becomes smaller when $H_{ext}$ increases [26]. In fact, similar qualitative explanation can be applied to FM [28], but it fails to justify the relation between $J_{crit}$ and $H_{ext}$. In addition, in the FM system, even using the macrospin model, $J_{crit}$ still decreases when $H_{ext}$ becomes larger. This demonstrates the critical differences between the SOT switching in Mn$_3$Sn and in perpendicular FM.

Furthermore, we find the SOT-induced switching polarities are opposite between Mn$_3$Sn and FM. It has been shown in Fig. 2(d) that the octupole will be switched to state a with $\varphi = 30°$, i.e., the z component of the octupole moment ($m_{oct,z}$) is larger than 0. We redraw the dynamics of $\mathbf{m}_{oct}$ in Fig. 4(a). In comparison, we study the switching of perpendicular FM in the same setup, i.e., the device consists of FM/HM bilayer where $\theta_{SH}>0$. When both $\mathbf{J}_c$ and $\mathbf{H}_{ext}$ are applied in +**y** direction, as shown in Fig. 4(b), the magnetic moment switches from +**z** to −**z**. This comparison shows that despite the similar device configuration, the underlying switching mechanism is different between Mn$_3$Sn and FM.

To complete the switching diagram, we further increase $H_{ext}$. Under a fixed $J_c=5\times10^{10}$ A/m$^2$, when $H_{ext}$ is increased above 275 Oe, the octupole moment will align with the magnetic field, i.e., $\varphi = 90°$ [see Fig. 5(a)]. Note that there is an abrupt change in the magnetic state when $H_{ext}$ exceeds the critical value ($H_{crit}$), dividing the phase diagram into two distinct regions, i.e., region 1 and 2 in Fig. 3(e). As shown in Fig. 5(b), for $H_{ext} > H_{crit}$, the switching results are independent of the initial state. When $\mathbf{J}_c$ is reversed, the result remains the same [see Fig. 5(c)]. This shows that SOT loses control on the magnetic state and only $\mathbf{H}_{ext}$ is effective. We therefore treat this as an unsuccessful SOT switching, which is illustrated in Fig. 3(e) as the light blue region.

In fact, the magnetization switching in configuration II has been experimentally observed recently [22]. However, the authors attributed it to the misalignment of crystalline orientation, and then explained the results using the switching mechanism in configuration I. Further experiments will be beneficial to clarify the underlying physics. Besides obtaining a high quality sample, we suggest a phase diagram measurement, including sweeping $\mathbf{H}_{ext}$ and $\mathbf{J}_c$, would be helpful for a better understanding.

During the preparation of this manuscript, the experimental study of switching Mn$_3$Sn in configuration II has been published [29], where the opposite switching polarity between Mn$_3$Sn and FM observed in our work was demonstrated experimentally, i.e., the handedness anomaly defined in their paper. However, to achieve the switching in their system, it requires the strain induced modification of exchange constant and anisotropy, especially the additional uniaxial anisotropy that pointing in the same direction for all the atoms. The additional anisotropy and the unequal exchange constant between sublattices may lead to the distortion of spin configuration [13] and further complicate the physical picture. In contrast, in our work, we propose a clean switching picture and demonstrate that these strain related modifications are not required, which can greatly simplify the device engineering.

**Conclusion**

We have demonstrated deterministic SOT switching of $Mn_3Sn$ in configuration II where the spin polarization is perpendicular to the kagome plane. We show that $J_{crit}$ is in the order of $10^{10}$ A/m$^2$ and $H_{ext}$ is around a hundred Oersted. We show that the switching result is uniquely determined by the combination of $\mathbf{J}_c$ and $\mathbf{H}_{ext}$. These properties are similar to the FM. However, the unique spin structure of $Mn_3Sn$ also leads to distinct features. The switching relies on the balance of effective SOT field and $\mathbf{H}_{ext}$, based on which we verify that $J_{crit}$ increases linearly with $H_{ext}$. In addition, the complex spin structure results in an opposite switching polarity compared to the FM. We conclude that the mechanism in FM cannot be directly applied to $Mn_3Sn$. Finally, we provide the switching phase diagram as a guideline for experimental demonstration.

†Corresponding Author: smeshis@buaa.edu.cn, zhuzhf@shanghaitech.edu.cn

The data that support the findings of this study are available from the corresponding author upon reasonable request.

**Acknowledgments**: This work was supported by National Key R&D Program of China (Grant No. 2022YFB4401700), National Natural Science Foundation of China (Grants No. 12104301 and 12104032), and Beijing Nova Program (Z211100002121123).

**References**

**Supplementary S1: Magnetization dynamics in configuration I**

The dynamics of octupole moment is shown in Fig. S1. As shown in Fig. S1(a), starting from state a, it can be switched to state f when $\mathbf{J}_c$ and $\mathbf{H}_{ext}$ are along $+\mathbf{x}$ and $-\mathbf{x}$ direction, respectively. Note that the $J_{crit}$ is very large to realize successful switching. Similarly, Fig. S1(d) shows that state c can be switched to state d, and Fig. S1(e) shows that state d can be switched to state c. However, as shown in Figs. S1(c) and S1(f), there is no switching under any combination of $\mathbf{J}_c$ and $\mathbf{H}_{ext}$ for state b and e. Interestingly, the effect of precession appears, which is absent in configuration II. We have ruled out the possibility of using an incorrect time step. The complete switching cases have been summarized in table S1. We have verified that the switching results of states c and d shown in table S1 are identical to that in [16]. To our knowledge, the results of the other four states have not been reported.

**Supplementary S2: Magnetization dynamics in configuration II under zero or small field**

As shown in Fig. S2(a), when $\mathbf{H}_{ext}$ is absent, the magnetic moments develop into oscillation. In addition, the oscillation frequency increases as the current is increased. When a small $\mathbf{H}_{ext} = 50$ Oe along the current direction is applied, the system remains in oscillation. Therefore, although [16] has studied this configuration, it might be the use of an inappropriate field strength that they did not obtain the deterministic switching. We have also shown that all the six stable states can be achieved by controlling the pulse width. As a result, this switching methodology is sensitive to the pulse width. In contrast, as shown in Fig. 2(d), in our system, the switching is independent of the pulse width, which is beneficial for the device application.

**Supplementary S3: Switching trajectories in configuration II**

Figure S3(a) shows the time evolution of $\mathbf{m}_{oct}$ for the switching from $\varphi = 210°$ to $\varphi = 30°$ for the same conditions in Fig. 2(d). It is found that the **x** component of $\mathbf{m}_{oct}$ is nearly zero during the switching process. Therefore, the magnetization switching is confined in the **y-z** plane. Figs. S3(b)-(e) shows the 3D trajectories of $\mathbf{m}_1$, $\mathbf{m}_2$, $\mathbf{m}_3$, and $\mathbf{m}_{oct}$. In contrast to the SOT switching in FM that exhibits precession during the magnetization reversal, $\mathbf{m}_{oct}$ is directly rotated to the minimum energy. It is clearly seen that the switching is precession less. The trajectory of the $\mathbf{m}_{oct}$ follows a highly regular arc during the switching, and its value remains constant. The cross-section view is summarized in Fig. S3(f). Furthermore, each atomic magnetic moment rotates from the initial state to the final state experiences the same angle with the $\mathbf{m}_{oct}$., i.e., 180° as shown in Fig. S3(f).

Figures

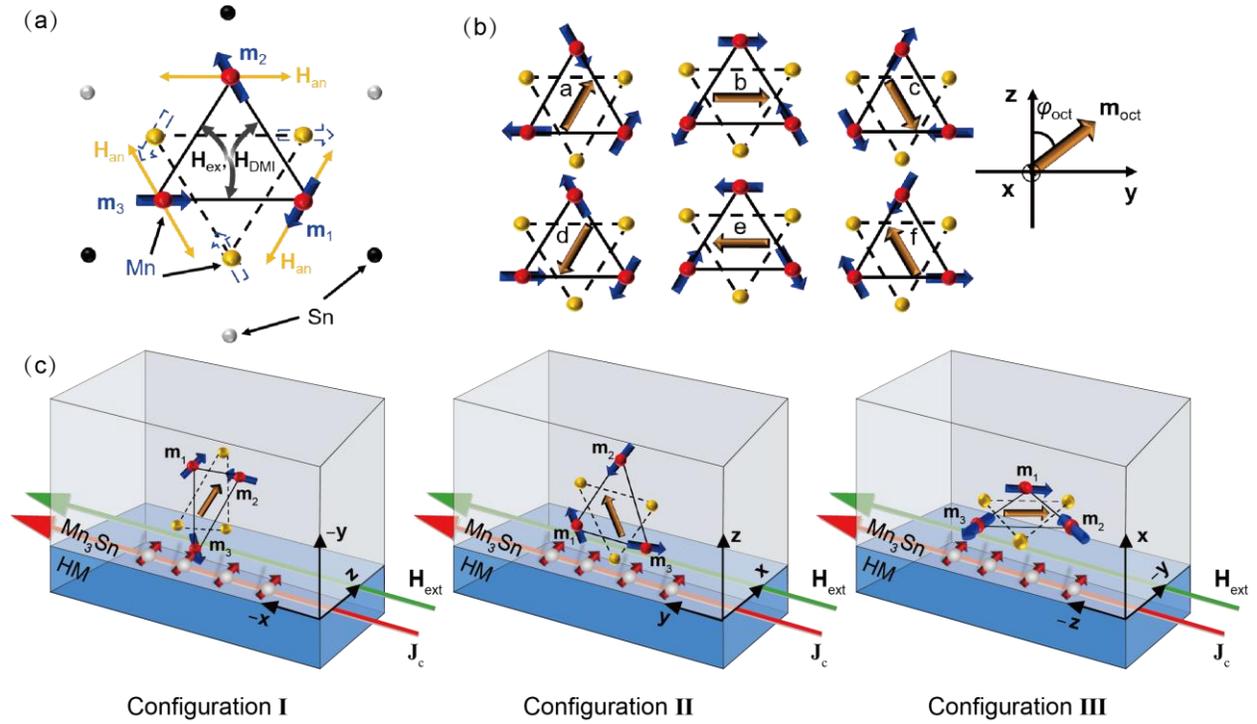

Fig. 1. (a) Atomic structure of $Mn_3Sn$. The red and yellow circles denote Mn atoms on the different layers. The black and silver circles denote Sn atoms on the same layer of yellow and red Mn, respectively. (b) The six equivalent states of $Mn_3Sn$, which are defined as state a, b, c, d, e, and f, respectively. The corresponding angles are $\varphi=30°, 90°, 150°, 210°, 270°, 330°$. (c) The three device configurations which are defined as configuration I, II, and III, respectively. In configuration I, the line connecting $m_1$ and $m_3$ points to +**y** direction.

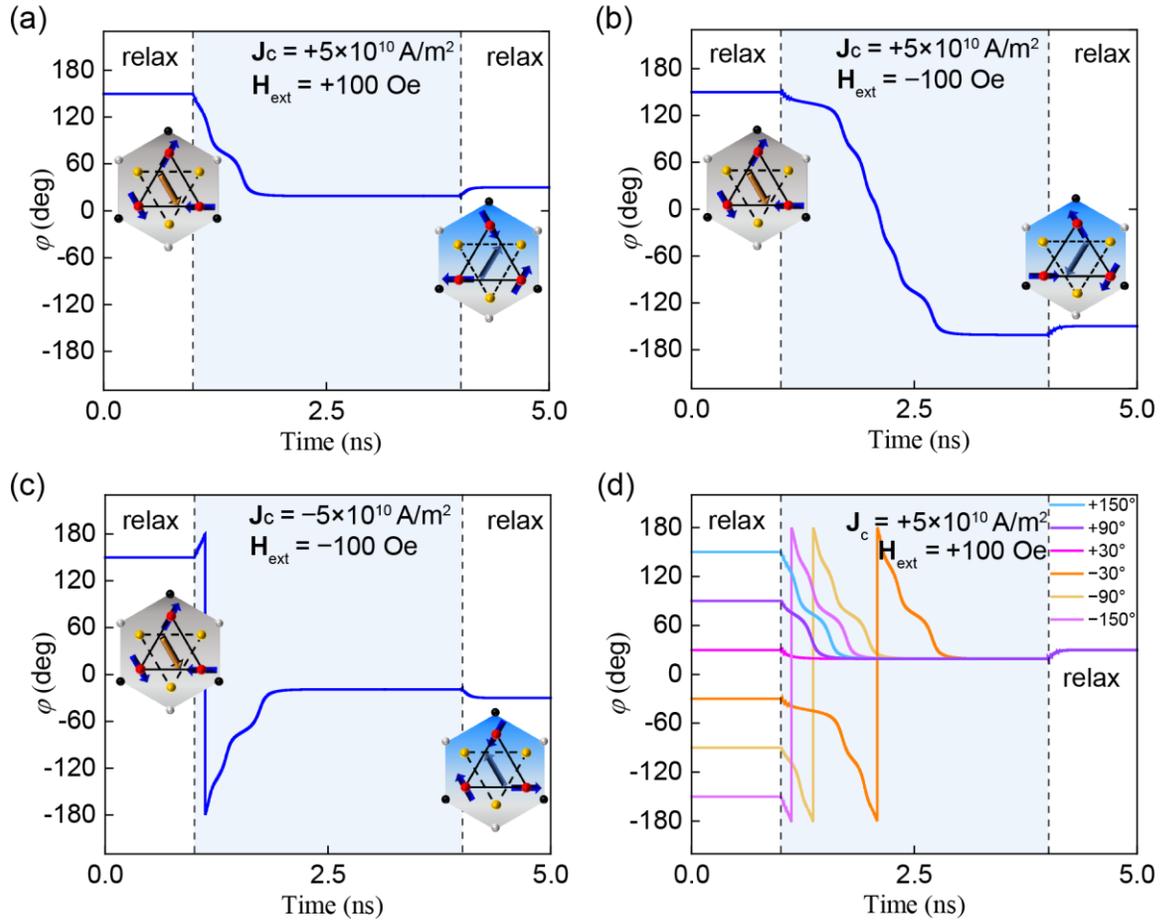

Fig. 2. The dynamics of octupole moment in configuration II. Before turning on $\mathbf{J}_c$ and $\mathbf{H}_{ext}$, the system is relaxed for 1 ns to reach the equilibrium state. At 4 ns, both $\mathbf{J}_c$ and $\mathbf{H}_{ext}$ are removed, and the system is relaxed to the new stable state. The insets show the spin structure for the initial and final states.

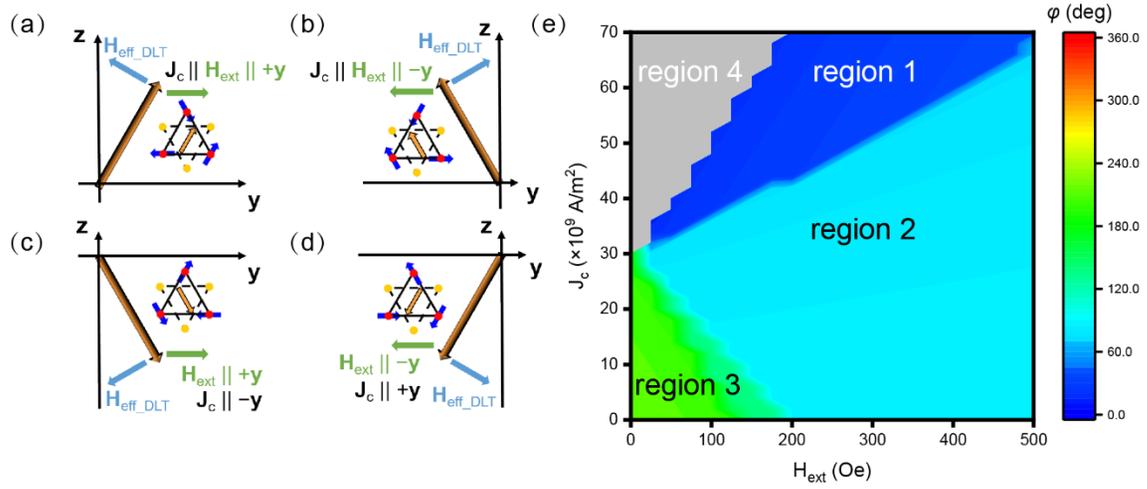

Fig. 3. (a)-(d) Illustration of effective fields from spin-orbit torque and external field. (e) The switching phase diagram when $J_c$ and $H_{ext}$ are varied. Region 1 denotes successful switching from state d to state a. The octupole moment aligns to $\mathbf{H}_{ext}$ in region 2. In region 3, the magnetic moments remain in the initial state. In region 4, the magnetic moments develop into oscillation.

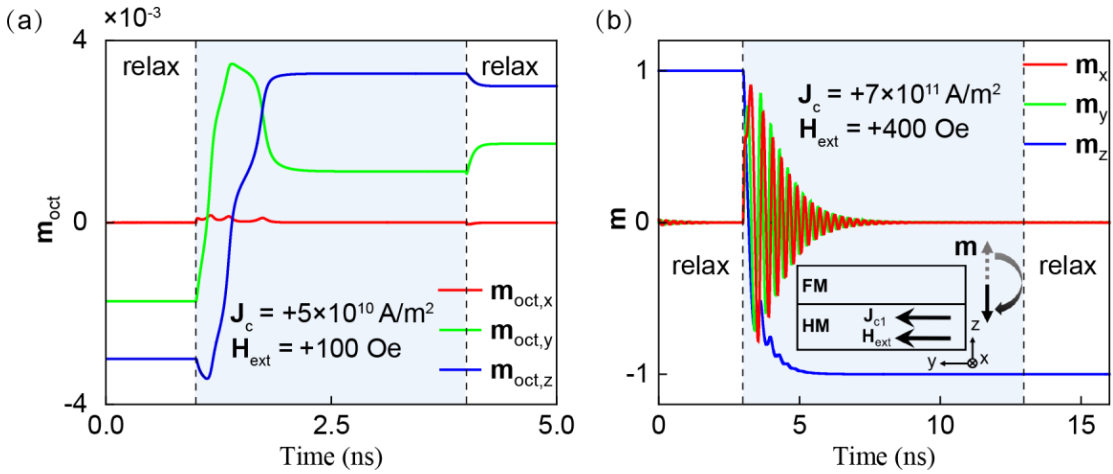

Fig. 4. Dynamics of the (a) octupole moment in $Mn_3Sn$ and (b) magnetization in FM. Both devices have $\theta_{SH} > 0$. $\mathbf{J}_c$ and $\mathbf{H}_{ext}$ in both cases are along +y direction.

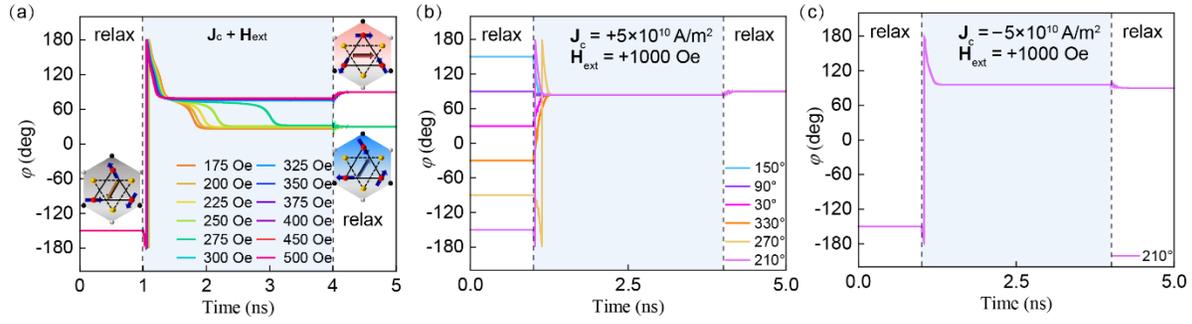

Fig. 5. (a) When $J_c=+5\times10^{10}$ A/m$^2$, the dynamics of octupole moment under different $H_{ext}$. Starting from different initial state, the dynamics of octupole moment when (b) $J_c=+5\times10^{10}$ A/m$^2$ and (c) $J_c=-5\times10^{10}$ A/m$^2$ with $H_{ext}=+1000$ Oe.

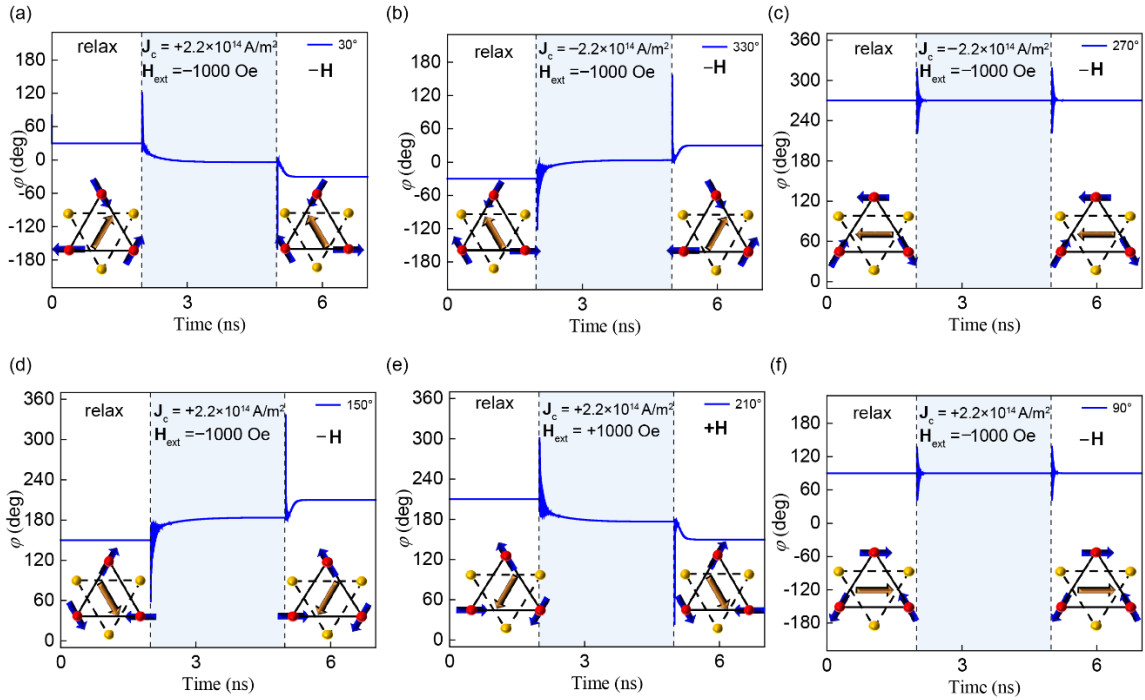

Fig. S1. In configuration I, the dynamics of octupole moment under different combination of $J_c$ and $H_{ext}$. The insets show the initial and final states.

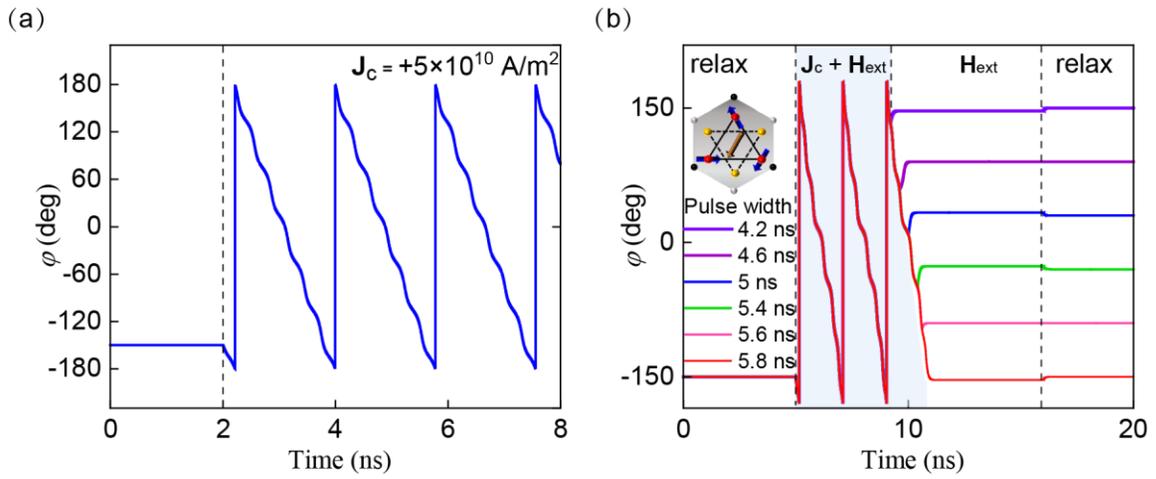

Fig. S2. In configuration II, oscillations of the octupole moment when (a) $\mathbf{H}_{ext} = 0$ Oe and (b) 50 Oe. Different pulse widths are applied in (b) that lead to different stable states.

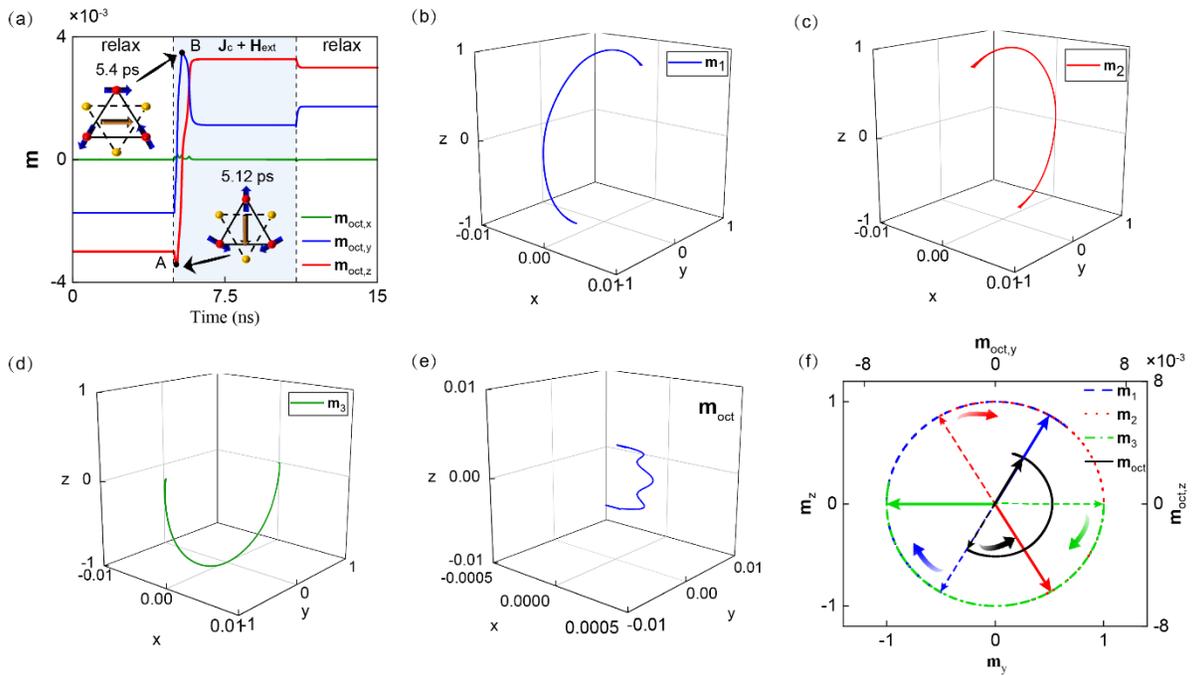

Fig. S3. (a) Dynamics of the octupole moment. (b)-(e) 3D trajectory of magnetic moments and the octupole moment. (f) Cross section view of the magnetic trajectory. The initial and final state are denoted by the dotted and solid arrow, respectively.

Table 1: Switching cases in configuration II.

| Initial State | $J_c=+y$, $H_{ext}=+y$ | $J_c=-y$, $H_{ext}=-y$ | $J_c=+y$, $H_{ext}=-y$ | $J_c=-y$, $H_{ext}=+y$ |
|---|---|---|---|---|
| a 30° | a 30° | f 330° | d 210° | c 150° |
| b 90° | a 30° | f 330° | d 210° | c 150° |
| c 150° | a 30° | f 330° | d 210° | c 150° |
| d 210° | a 30° | f 330° | d 210° | c 150° |
| e 270° | a 30° | f 330° | d 210° | c 150° |
| f 330° | a 30° | f 330° | d 210° | c 150° |

Table S1: Switching cases in configuration I.

| Initial State | $J_c=-x$, $H_{ext}=-x$ | $J_c=+x$, $H_{ext}=+x$ | $J_c=-x$, $H_{ext}=+x$ | $J_c=+x$, $H_{ext}=-x$ |
|---|---|---|---|---|
| f 330° | a 30° | a 30° | f 330° | f 330° |
| e 270° | e 270° | e 270° | e 270° | e 270° |
| d 210° | c 150° | c 150° | d 210° | d 210° |
| c 150° | c 150° | c 150° | d 210° | d 210° |
| b 90° | b 90° | b 90° | b 90° | b 90° |
| a 30° | a 30° | a 30° | f 330° | f 330° |